\PassOptionsToPackage{svgnames,table,dvipsnames}{xcolor}
\documentclass[journal]{vgtc}                     

\onlineid{0}

\vgtccategory{Research}

\vgtcpapertype{please specify}

\title{Radiance Fields in XR: A Survey on How Radiance Fields\\are Envisioned and Addressed for XR Research \thanks{© 2025 IEEE.  Personal use of this material is permitted.  Permission from IEEE must be obtained for all other uses, in any current or future media, including reprinting/republishing this material for advertising or promotional purposes, creating new collective works, for resale or redistribution to servers or lists, or reuse of any copyrighted component of this work in other works.}}

\author{%
  \authororcid{Ke Li}{0000-0003-0828-029X},
  \authororcid{Mana Masuda}{0000-0002-9050-5306},
  \authororcid{Susanne Schmidt}{0000-0002-8162-7694},
  \authororcid{Shohei Mori}{0000-0003-0540-7312}
}

\authorfooter{
  \item
  	Ke Li is with University of Hamburg, Germany.  \\ E-mail: ke.li@uni-hamburg.de
  \item
  	Mana Masuda is with Keio University, Japan. \\
  	E-mail: mana.smile@keio.jp

  \item Susanne Schmidt is with University of Canterbury, New Zealand.  \\ E-mail: susanne.schmidt@canterbury.ac.nz

  \item Shohei Mori is with University of Stuttgart, Germany. \\ E-mail: s.mori.jp@ieee.org
}

\abstract{
    The development of radiance fields (RF), such as 3D Gaussian Splatting (3DGS) and Neural Radiance Fields (NeRF), has revolutionized interactive photorealistic view synthesis and presents enormous opportunities for XR research and applications. However, despite the exponential growth of RF research, RF-related contributions to the XR community remain sparse. To better understand this research gap, we performed a systematic survey of current RF literature to analyze (i) how RF is envisioned for XR applications, (ii) how they have already been implemented, and (iii) the remaining research gaps. We collected $\mathbf{365}$ RF contributions related to XR from computer vision, computer graphics, robotics, multimedia, human-computer interaction, and XR communities, seeking to answer the above research questions. Among the $\mathbf{365}$ papers, we performed an analysis of $\mathbf{66}$ papers that already addressed a detailed aspect of RF research for XR. With this survey, we extended and positioned XR-specific RF research topics in the broader RF research field and provide a helpful resource for the XR community to navigate within the rapid development of RF research. 
} 


\teaser{
  \centering
  \includegraphics[width=\linewidth]{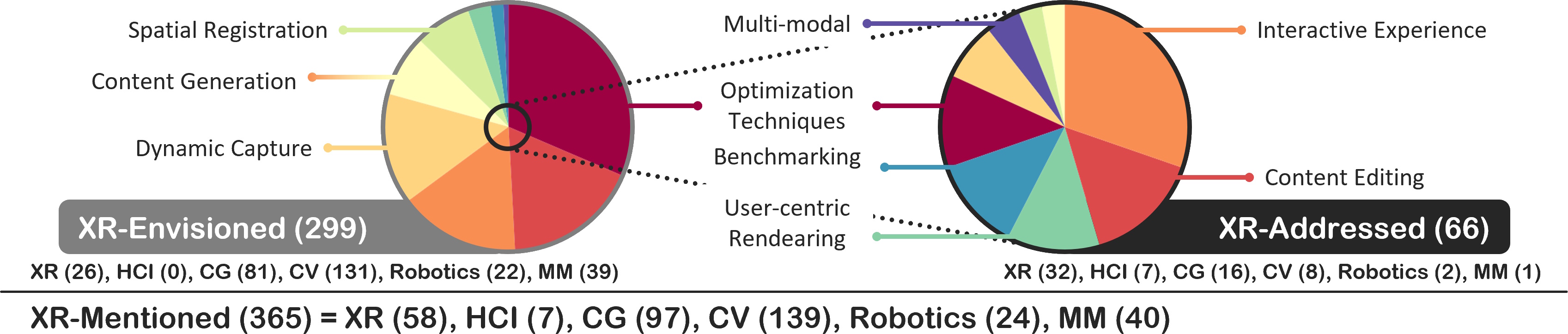}
  \caption{
  We observe a gap between the anticipated interests in RF-related technologies expected to advance new XR applications (left) and the actual interests that are considered the main factors in XR (right).
  This survey paper investigates and discusses how RF technologies, including NeRF and 3DGS, are envisioned for XR (i.e., \faLightbulb~XR-Envisioned) and how they are actually implemented for XR (i.e., \faClipboardCheck~XR-Addressed) in scientific papers from XR, HCI, CV, CG, Robotics, and MM communities via systematic review.%
  }
  \label{fig:teaser}
}

\graphicspath{{figs/}{figures/}{pictures/}{images/}{./}} 
\usepackage{xcolor} 

\usepackage{pdfpages}
\definecolor{NavyBlue}{named}{NavyBlue}

\usepackage{graphicx}
\usepackage{tabu}                      
\usepackage{booktabs}                  
\usepackage{lipsum}                    
\usepackage{mwe}                       

\usepackage{mathptmx}                  

\usepackage{fontawesome5}
\usepackage{amsmath, amssymb}
\usepackage{enumitem}

\begin{document}

\firstsection{Introduction}

\maketitle
 
Machine learning revolutionized how we optimize scene representations for given images.
Radiance fields (RF) \cite{Mildenhall2020NeRF,Kerbl20233DGS} are noticeable examples that have enabled efficient encoding and rendering of complex scene appearance since 2020.
This innovation introduced significant potential for XR applications such as remote collaboration \cite{reynolds2024pop}, teleoperation \cite{li2024reality}, and telepresence \cite{avatar_texel_aligned}
and could potentially replace the traditional meshes, point clouds, and voxels.
Advances in efficient training and rendering of RF \cite{Mller2022InstantNG,Kerbl20233DGS} further expanded the opportunities in XR applications that require high-resolution and real-time rendering.
RF is considered to provide a novel technical foundation for enhancing immersive video see-through head-mounted displays \cite{VST_neural_passthrough}, creating dynamic photorealistic avatars for the metaverse \cite{avatar_texel_aligned}, and low-cost generation of digital twins for real-world XR applications \cite{li2024reality}. 
 
Due to its significance, there has been an exponential growth of RF research in computer vision (CV), computer graphics (CG), robotics, and multimedia (MM) communities since the first introduction of the neural radiance field (NeRF) paper \cite{Mildenhall2020NeRF}. While many of these technical papers envisioned enormous impacts of RF technology for XR, the actual implementation of RF for XR-focused research and applications remains comparably sparse.
For instance, \figurename~\ref{fig:teaser} shows that 203 RF-related papers were published at the Conference on Computer Vision and Pattern Recognition (CVPR) in 2024, with 68 explicitly mentioning XR as a target application domain. However, in the proceedings of leading XR conferences — IEEE VR 2024 and IEEE ISMAR 2024 — only 11 contributions (including 4 conference papers and 7 extended abstracts) were related to RF. Among these, just 5 directly addressed an XR-related RF research question through technical benchmarks or user studies conducted in actual XR settings. This indicates a significant research gap between the community’s initial vision for RF in XR and its current integration into XR research and applications.


To better understand this research gap, this paper presents a systematic survey on how radiance fields are envisioned for XR and how they have been implemented. While numerous surveys on radiance fields exist across various domains — including robotics \cite{NeRF_robot_survey} and 3D computer vision  \cite{NeRF_3DV_Survey} — to the best of our knowledge, no dedicated survey has been conducted on radiance fields for XR. Moreover, other existing general radiance field surveys typically only provide a broad overview of current research \cite{TVCG_2024_3DGS_Survey}, offering limited insights into crucial XR-specific topics such as foveated rendering, haptic rendering, telepresence, 3D interaction, and perception. As a result, the XR community currently has limited resources in evaluating the advancement of radiance field technology for XR and assessing its applicability in real-world immersive XR applications. This survey paper aims to bridge such research gaps by investigating the following research questions:
\begin{itemize}[left=0pt]
    \item \textbf{RQ1}: How are radiance fields envisioned for XR?
    \item \textbf{RQ2}: What aspects of radiance fields have been implemented in the current literature, specifically, through technical benchmarks or user studies in actual immersive XR settings?
    \item \textbf{RQ3}: What are the remaining research gaps for widespread adaptation of radiance fields in XR?
\end{itemize}

To address these research questions, we conduct a systematic survey of key RF-related literature across the CV, CG, robotics, MM, human-computer interaction (HCI), and XR communities, drawing from the Association for Computing Machinery (ACM) and Institute of Electrical and Electronics Engineers (IEEE) databases.
From the initial screening of 1304 papers, we identify 365 as ``\textit{XR-Mentioned}'' papers — those that mention XR as a potential application — among which we labeled \textbf{66} as ``\textit{XR-Addressed}'' papers — those that contribute to RF research through technical benchmarks or user experiments in immersive XR settings. To answer \textbf{RQ1}, we perform a thematic analysis of the ``\textit{XR-Mentioned}'' corpus, identifying \textbf{nine} key themes that reveal how RF is envisioned for XR. For \textbf{RQ2}, we conduct an in-depth literature review of the ``\textit{XR-Addressed}'' papers, analyzing state-of-the-art RF research for XR topics, including foveated rendering, haptic rendering, remote collaboration, and 3D interaction. To address \textbf{RQ3}, we compare the findings from our thematic analysis and literature review to identify key challenges and future research opportunities for RF in XR. 

%

\begin{figure}[t]
    \centering
    \includegraphics[width=\columnwidth]{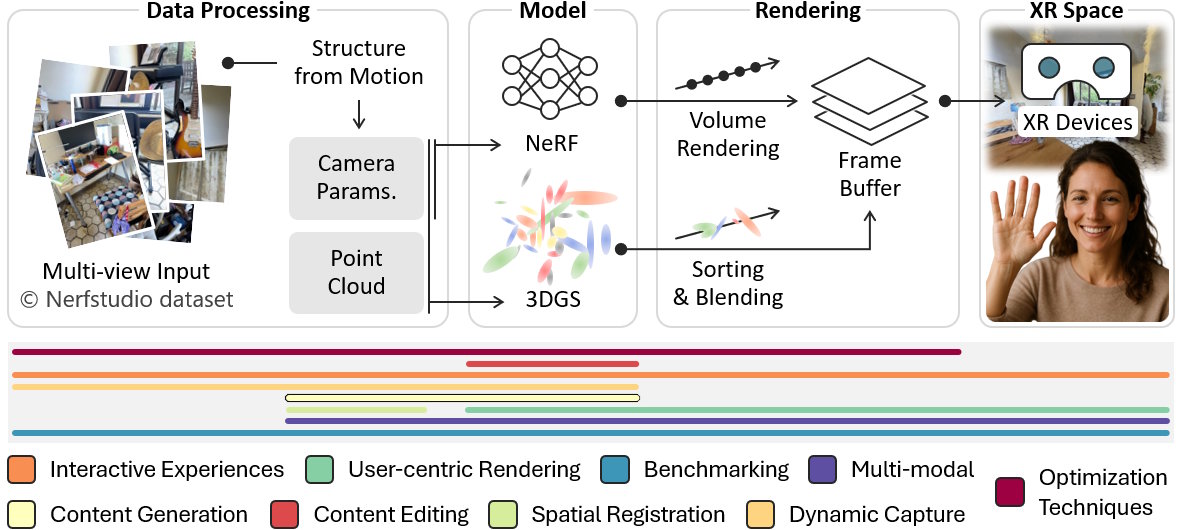}
    \caption{RF pipeline for XR. Through analysis by synthesis, a scene model (i.e., NeRF, 3DGS, or their variants) is optimized to match rendered and input views.
    XR applications utilize the rendering technology for new ``Interactive Experiences,''
    ``User-centric Rendering'' for high-performance stereo rendering,
    ``Benchmarking'' XR systems,
    extending the concept to ``Multi-modal'' I/O,
    plausible ``Content Generation,''
    interactive ``Content Editing,''
    ``Spatial Registration'' such as SLAM,
    extending architecture for ``Dynamic Capture,''
    or ``Optimization Techniques'' for demanding XR scenarios.
    Different color bars indicate ranges of individual topics over the RF pipeline.}
    \label{fig:basics}
\end{figure}





\section{Background} \label{sect:terms}

\noindent
\textbf{XR:}
Many definitions exist to cover different ranges of reality technologies, including Virtual Reality (VR) \cite{sutherland1965ultimate}, Augmented Reality (AR) \cite{azuma1997survey}, Mixed Reality (MR) \cite{milgram1994taxonomy,skarbez2021revisiting}, eXtended Reality (xR) \cite{mann2018all}, Diminished Reality (DR) \cite{mori2017survey}, and Mediated Reality \cite{mann1994mediated}. Each term discusses which types of realities should be included or excluded, thereby defining a new domain. To avoid any conflicts and to be inclusive, we use the term XR, where X can represent any kind of reality. 



\noindent
\textbf{Radiance Fields:} RF is a class of photorealistic scene representations that model the spatial and directional distribution of light in a 3D environment. These representations encode the color and density of a scene in various forms of data structures, such as neural networks or 3D Gaussians, enabling realistic view synthesis and high-fidelity 3D reconstructions. NeRF and 3DGS are two prominent approaches within this category, differing primarily in their underlying data structures and rendering techniques.
A common integration strategy involves rendering RF into a frame buffer for XR displays (\figurename~\ref{fig:basics}). This keeps the original renderer untouched by feeding view frustum information to and receiving the results from the renderer in XR middleware (e.g., Unity). 
3DGS, as an explicit data representation, offers greater potential for tighter integration by leveraging shader codes for optimal performance.
When depth rendering is available, on-screen composition with other frame buffers, such as those used for meshes, is possible. For a detailed comparison of 3DGS, NeRF, and textured meshes for XR applications, we refer readers to the supplementary materials. 

\noindent
\textbf{Neural Radiance Fields:}
NeRF represents a continuous 3D scene as parameters of a multi-layer perceptron (MLP). The MLP is trained using a set of multi-viewpoint images and their camera poses typically obtained from a structured from motion (SfM) \cite{structure_from_motion} algorithm (\figurename~\ref{fig:basics}).
For any given 3D point and 2D viewing direction, a NeRF model infers the corresponding color and density values \cite{Mildenhall2020NeRF}.
Through volume rendering, color and density samples are accumulated per ray to represent complex scene geometry and photometric appearance.
While the original NeRF model does have limitations in performance, recent advances in efficient neural network architecture and sampling strategies have significantly improved the performance \cite{Mller2022InstantNG}. 
We refer NeRF variants to NeRF with different data structures, such as k-planes \cite{KPlanes_NeRF} and instant neural primitives \cite{Mller2022InstantNG}, as they share general ideas (i.e., neural implicit scene representation). For further technical details, we refer to a previous survey focusing technical aspect of  NeRF\cite{NeRF_3DV_Survey}.

\noindent
\textbf{3D Gaussian Splatting:}
3DGS \cite{Kerbl20233DGS} is an explicit 3D scene representation that models appearances using a set of anisotropic 3D Gaussians. Similarly to NeRF, it uses calibrated camera parameters and also a sparse point cloud from SfM to initialize 3D Gaussians (\figurename~\ref{fig:basics}). The method employs a rasterization approach by projecting the 3D Gaussians onto 2D, resulting in a GPU-friendly differential rendering technique that achieves faster rendering speeds compared to typical NeRF-based implicit 3D scene representations. For further technical details, we refer to a previous survey focusing on 3DGS and its optimization \cite{TVCG_2024_3DGS_Survey}.

\section{Methodology} \label{sect:methodology}

This section outlines the systematic survey process following PRISMA 2020 guidelines \cite{PRISMA2020}. A detailed PRISMA flowchart is provided in the supplementary material.

\subsection{Data Acquisition: Collecting \textit{RF-Mentioned} Corpus}
To ensure coverage of a wide range of RF-related research conferences, journals, and venues, we started our search process using one of the largest computer science online databases ACM Digital Library (ACM DL) and the IEEE Xplore library (Search date: February 26, 2025). The year range starts from March 2020, when the first NeRF paper is published, and terminates on the search date.  Our search queries cover typical terms and definitions of RF-related concepts, their variants, and abbreviations in all metadata of the paper entry in the database: \emph{``neural radiance field'' OR ``radiance field'' OR ``3D gaussian splatting'' OR ``gaussian splatting'' OR ``3DGS'' OR ``NeRF''}. The initial search results in a total of $\mathbf{2,259}$ items, with $\mathbf{889}$ from ACM and $\textbf{1,676}$ from IEEE. In addition, despite it's recency and due to its high relevance,  we also include the conference proceedings of IEEE VR 2025, which contains an additional 17 contributions.  

To ensure the inclusion of only high-quality contributions, we selected core conferences and journals from key RF-related research communities. These include core XR venues: ACM Virtual Reality Software and Technology (VRST) ($n=4$), IEEE VR ($n=23$), IEEE ISMAR ($n=8$), IEEE Transactions on Visualization and Computer Graphics (TVCG) ($n=48$),; core CG venues and journals: ACM Transactions on Graphics (TOG) ($n=168$), ACM SIGGRAPH and SIGGRAPH Asia ($n=243$); core HCI venues: ACM SIGCHI ($n=9$) and ACM UIST ($n=6$); core robotics venues: the IEEE/RSJ International Conference on Intelligent Robots and Systems (IROS) ($n=59$), the IEEE International Conference on Robotics and Automation (ICRA) ($n=43$); core CV venues: IEEE/CVF International Conference on Computer Vision (ICCV) ($n=143$), IEEE/CVF Conference on Computer Vision and Pattern Recognition (CVPR) ($n=419$); and multimedia venue: the ACM Multimedia Conference (MM) ($n=132$). All selected conferences and journals hold a CORE ranking of A or A*. 
For the core XR and core HCI venues, we also included the extended abstract proceedings (poster and research demo) to ensure that emerging research topics in the XR community that might still be in the stage of preliminary work are included. Further, we removed $\mathbf{37}$ items which are either duplicate results from the ACM DL, or irrelevant results ( for example, using the physical NeRF gun for user studies rather than radiance fields). This results in a total of $\mathbf{1,305}$ \textit{RF-Mentioned} papers for further screening and selection. 


\subsection{Screening and Inclusion Criteria: Determine \textit{XR-Mentioned} and \textit{XR-Addressed}}


Our main goal in the literature screening step is to identify all \textit{XR-Mentioned} contributions within the \textit{RF-Mentioned} corpus. For the initial screening, we employed an automated screening technique using a Python script. If a paper contains any of the following filtering keywords in the main text (excluding references), the paper is labeled as \textit{XR-Mentioned}: 

\vspace{0.5em}
\noindent
\textbf{XR filtering keyword list}: \textit{virtual reality}, \textit{augmented reality}, \textit{metaverse}, \textit{extended reality}, \textit{diminished reality}, \textit{augmented virtuality}, \textit{mixed reality}, \textit{immersive}, \textit{telepresence}, \textit{holoportation}. 
\vspace{0.5em}

This initial screening process excluded $\mathbf{846}$ items, with the remaining $\mathbf{459}$ papers eligible for further manual screening and selection by the authors of this paper to determine which contributions have specifically addressed an aspect of XR.
The main criteria for inclusion in the \textit{XR-Addressed} categories are whether the contribution has an actual technical benchmark in an actual XR setting (e.g. a VR head mounted display (HMD) or interactive AR/MR applications on conventional 2D display setups), whether the contribution shows clear potentials to transfer their results into XR applications, or if user-centric studies were performed involving testing with human subjects in XR.
Borderline contributions are highlighted (e.g., contributions that developed 360 rendering techniques but didn't evaluate it using an HMD) and separately discussed by all the authors altogether to determine if they should be included in the \textit{XR-Addressed} category.  
This screening step results in the following categorizations: 
\begin{itemize}[left=0pt]
    \item \faClipboardCheck~\textbf{XR-Addressed ($n=66$)}: RF contributions addressing XR-specific aspects of it. They are included for the final in-depth literature review. 
    \item \faLightbulb~\textbf{XR-Envisioned ($n=299$)}: RF contributions mentioned XR keywords only a few times, but no validations in an actual XR setting (e.g., rendering in XR settings or performing user-centric evaluations). We extracted the text from each paper that included XR keywords. We included these texts in a further thematic analysis to identify the key themes/taxonomy categories in how the community has envisioned RF for XR research and applications. 
    \item \textbf{Others ($n=94$)}: Items are further manually filtered and excluded as they only mention RF-related keywords but contain no major contributions to RF (e.g., papers focusing on signed distance field (SDF) rather than NeRF/3DGS, or papers that mentioned ``immersive'' referring to video games rather than XR settings).   
\end{itemize}

We further categorized \faClipboardCheck~papers into the following:
\begin{itemize}[left=0pt]
    \item \faUserPlus~XR-Study (n=24): contributions that include an in-depth evaluation with users via XR setups and scenarios (e.g., performance evaluation using an XR headset) or contributions that include technical benchmark using XR-specific metrics whose results are directly transferrable to XR applications (e.g., performance tests of a foveated rendering algorithm using pre-collected user eye-gaze data).
    \item \faUserMinus~XR-Showcase (n=42): contributions that mostly showcase XR applications in the form of preliminary prototypes or demonstrations. Some of these do not have specific user-involved benchmarking or evaluations in the firm context of XR, but we value these researches, whose findings could easily transfer to actual immersive XR settings. 
\end{itemize}

\subsection{Iterative Taxonomy Development Process}

Based on the identified XR-envisioned (n=299) and XR-addressed contributions (n=66), a total of 365 contributions were eligible for full-text screening and iterative taxonomy development.


\textbf{Characteristics and dimension identification}:  In the first step of the taxonomy development process, each author was assigned a set of papers for full-text screening to identify the main contributions. After screening, the authors annotated each paper with 2-3 keywords summarizing the key contributions. Based on these keywords, we grouped the most frequently occurring ones into several main themes while keeping in mind of their relevance to XR while developing the main themes. After several iterative discussions, one of the authors compiled the remaining keywords and themes to formulate the final categorizations and taxonomy for XR. \figurename~\ref{fig:basics} presents the final taxonomy with different categories of RF research captured within nine main research themes: 
\textit{optimization techniques},
\textit{dynamic capture},
\textit{spatial registration},
\textit{content generation},
\textit{content editing},
\textit{multi-modal},
\textit{benchmarking},
\textit{user-centric rendering},
\textit{interactive experiences}. The taxonomy is structured so that the themes become increasingly relevant to interactivity and human factors in XR experiences from left to right.


\textbf{Tagging:} Using the developed taxonomy, we assigned each of the 365 contributions to one of the nine key RF research themes. For the 299 XR-mentioned papers, each paper was allocated to one of the co-authors for tagging. In cases where a contribution could fit into multiple categories, we identified the primary contribution and assigned the paper to the most appropriate category. Edge cases and borderline papers were flagged and discussed collectively among all co-authors in meetings, where tagging decisions were made through voting. For the 66 XR-addressed contributions, two authors independently tagged each paper to ensure coherence and accuracy. In instances where the two authors disagreed, the paper was tabled for discussion among the entire group of co-authors to reach a consensus.
    




\section{Gaps in Trends} \label{sect:trends}

Understanding which research communities contribute to XR-relevant RF development — and how their efforts differ between envisioning (\faLightbulb) and implementing XR use cases (\faClipboardCheck) — is key to identifying gaps, guiding collaborations, and accelerating progress in the field. In this section, we analyze publication trends across communities and reflect on their implications for XR research.

\begin{figure}[t]
    \centering
    \includegraphics[width=\columnwidth]{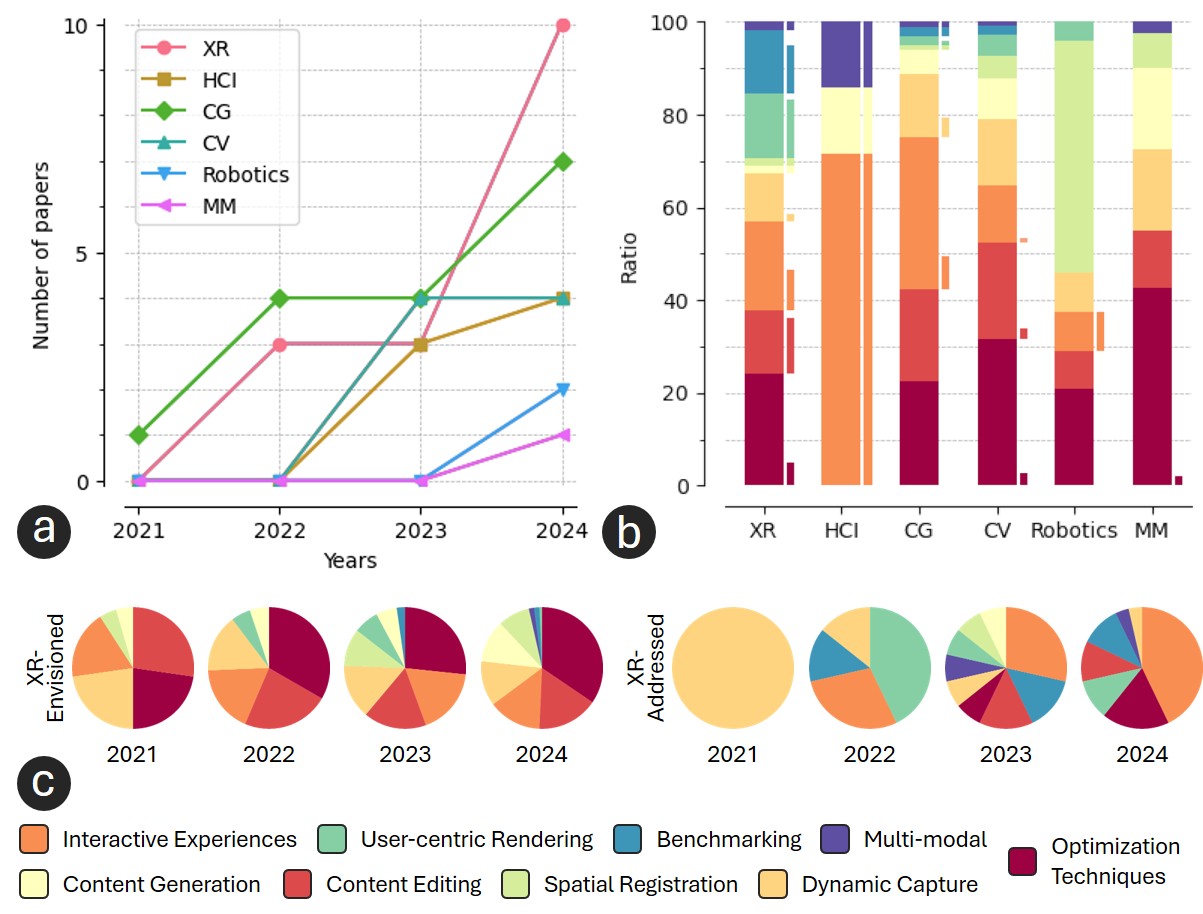}
    \caption{Trends in XR-related RF papers.
    (a) Growth of XR-Addressed papers over the years in the communities ('21--'24).
    (b) Interest populations in XR-Mentioned papers (2021--until IEEE VR '25 conference). The smaller bars indicate XR-Addressed papers.
    (c) Interest populations over the years ('21--'24).
    }
    \label{fig:trends}
\end{figure}

\textbf{Communities and Interests:}
A growing interest in XR is observed in RF research (\figurename~\ref{fig:trends}a).
Notably, in 2024, the XR community published more XR-Addressed papers than other communities. The XR and HCI communities are the main contributors to \textit{XR-Addressed} (\faClipboardCheck), whereas CV, CG, Robotics, and MM communities typically mention XR only as a benchmarking or motivation.

We observe varied balances in XR-Envisioned (\faLightbulb) and XR-Addressed (\faClipboardCheck) papers across communities:
XR (\faLightbulb: 26, \faClipboardCheck: 32),
HCI (\faLightbulb: 0, \faClipboardCheck: 7),
CG (\faLightbulb: 81, \faClipboardCheck: 16),
CV (\faLightbulb: 131, \faClipboardCheck: 8),
Robotics (\faLightbulb: 22, \faClipboardCheck: 2),
and MM (\faLightbulb: 39, \faClipboardCheck: 1).
While the XR community published a marginal number of papers ($n=57$), it contributed the most \faClipboardCheck papers, surpassing both CV ($n=139$) and CG ($n=97$). In contrast, CG, CV, Robotics, and MM communities contribute primarily to \faLightbulb papers, which often discuss XR as a future application area or aspirational use case.
The HCI community publishes only \faClipboardCheck papers, though in limited numbers. These distinctions underscore the need to examine more closely why certain communities contribute fewer system-level or human-in-the-loop XR works, despite frequent mentions of XR potential.

Importantly, we caution against interpreting XR mentions as superficial or misleading. Rather, they often reflect future aspirations or alignment with long-term XR goals. Many of these papers propose technically advanced methods and explicitly mention XR as a target domain. In this light, non-XR communities may play a key role by envisioning how their technical advances can support XR systems despite not tackling full-system integration or evaluation.

\figurename~\ref{fig:trends}b shows the overall interest of each community in applying RF to XR across nine topic areas. The XR, CG, and CV communities show interest in all nine topics, whereas HCI, Robotics, and MM focus on fewer areas (three, six, and six topics, respectively).
However, only the XR community addresses all nine topics in implementation. CG and CV address six and three topics, respectively, while the HCI community focuses implementation efforts on three. Notably, the Robotics community demonstrates interest in Spatial Registration ($n=12$), but actual implementations appear only in Interactive Experiences ($n=2$). Similarly, the MM community has a high number of envisioned papers in Optimization Techniques for XR ($n=39$), but only one addresses XR in practice.

This asymmetry reflects a broader challenge: XR researchers must often rely on external technical developments without direct evidence of those technologies' suitability in XR contexts. This reliance complicates the selection and adaptation of RF methods for XR systems. Conversely, implementing and evaluating RF methods within XR contexts can uncover new, system-level insights that are not evident from standalone evaluations.

\textbf{Envisioned vs. Addressed Topics:}
XR has been envisioned (\faLightbulb) across all nine topics:
Multi-modal ($n=2$),
Benchmarking ($n=5$),
User-centric Rendering ($n=9$),
Spatial Registration ($n=22$),
Content Generation ($n=24$),
Dynamic Capture ($n=43$),
Interactive Experiences ($n=47$),
Content Editing ($n=53$), and
Optimization Techniques ($n=94$).
XR has also been addressed (\faClipboardCheck) in all nine topics:
Content Generation ($n=2$),
Spatial Registration ($n=2$),
Multi-modal ($n=3$),
Dynamic Capture ($n=5$),
User-centric Rendering ($n=8$),
Optimization Techniques ($n=8$),
Benchmarking ($n=8$),
Content Editing ($n=10$), and
Interactive Experiences ($n=20$).
XR-Mentioned and XR-Addressed papers populate different topics (\figurename~\ref{fig:teaser}).

Envisioned papers frequently frame XR relevance through single-sentence mentions, often placed in the introduction (e.g., ``Novel view synthesis and real-time rendering from monocular video are critical not only for 3D vision tasks but also for practical applications such as virtual reality and video games,'' \cite{shao2025frequency}) or conclusion (e.g., ``This application can be adapted and used in many circumstances of virtual reality and augmented reality, e.g., telepresence and sports broadcasting,'' \cite{wang2022nerfcap}). Many use XR as a motivating lens to justify further technical performance improvements.

Over the last four years, such envisioning has appeared consistently in Optimization Techniques, Interactive Experience, Content Editing, and Dynamic Capture (\figurename~\ref{fig:trends}c). However, actual implementations have emerged in a more scattered and less predictable fashion.
These differences between envisioned potential and realized implementation raise questions about the bottlenecks researchers face when transitioning from algorithm design to working XR systems.

\textbf{Open Source:}
To accelerate progress and mitigate replication crises in XR research, open-source software (OSS) is essential \cite{swan2023replicationcrisis}. We surveyed available OSS projects associated with XR-Addressed and XR-Envisioned papers as of March 31, 2025, excluding empty or placeholder repositories.
Among XR-Addressed papers, $34.8\%$ ($n=23$ out of 66) released OSS (XR: $25.0\%=8/32$, HCI: $14.3\%=1/7$, CG: $50.0\%=8/16$, CV: $50.0\%=4/8$, Robotics: $100.0\%=2/2$, MM: $0.0\%=0/1$).
Among XR-Envisioned papers, $62.5\%$ ($n=112$ out of 187) released OSS (XR: $46.2\%=12/26$, HCI: no XR-Envisioned papers, CG: $61.7\%=50/81$, CV: $76.3\%=100/131$, Robotics: $45\%=10/22$, MM: $38,5\%=15/39$).
These OSS release rates appear to reflect disciplinary norms, institutional IP policies, and authors’ strategic choices. However, for XR-Addressed research, the relatively low availability of OSS can hinder reproducibility and slow community-wide validation and adoption.
Improving OSS availability, particularly for system-level implementations, will be crucial for strengthening the credibility, usability, and future impact of RF research in XR.

\section{Radiance Fields for XR} \label{sect:taxonomy}
We discuss the \textit{XR-Addressed} contributions (Tables \ref{XR-study} and \ref{XR-Showcase}) within each of the nine key themes identified through the iterative taxonomy development process described in Section \ref{sect:methodology}. We highlight the current achievements and identify research gaps and opportunities within each theme and category.

\subsection{Interactive Experiences}
The interactive experiences theme covers a diverse range of topics related to interactive XR experiences using RF representations, such as collaborative systems, photorealistic 3D virtual humans, and RF-enabled industrial and medical applications.

\textbf{Collaborative systems:} Three of the considered papers explore the potential of RF for asymmetric XR collaboration, where a local user captures their real-world environment, e.g. with a 360 camera \faUserPlus \cite{huang2024virtualnexus}, a mobile phone LiDAR scanner \faUserMinus \cite{reynolds2024pop}, or a head-mounted RGBD camera \faUserPlus \cite{sakashita2024sharednerf}, and sends the model to a remote user.
The remote user can then inspect the virtual replica of the local user's environment using immersive technology \cite{huang2024virtualnexus} or a desktop PC \cite{reynolds2024pop, sakashita2024sharednerf}.
Both Huang et al. \cite{huang2024virtualnexus} and Reynolds et al. \cite{reynolds2024pop} additionally provide the remote user with virtual annotation tools, with the results being embedded in the local user's view using AR technology.
While the \textit{Thing2Reality} platform by Hu et al. is also presented in a telepresence context, it mainly focuses on the transition of virtual objects between a shared 2D and 3D space \faUserMinus \cite{hu2024experiencing}. Using conditioned diffusion models, multiple views of a 2D object are generated and used as input to 3DGS to create a 3D representation of the pictured object.
The \textit{RealityGit} system by Li et al. highlights version control as a particular aspect of collaborative work \faUserMinus \cite{li2023realitygit}. \textit{RealityGit} allows remote users to asynchronously visualize and edit the current and historical states of the local NeRF environment and to suggest changes to the local user.

\textbf{Avatars and agents:} The largest portion of papers addressing XR for \textit{Interactive Experiences} investigates the use of RF to create body representations of virtual characters, controlled either by the user (i.e. avatars) or by the computer (i.e. agents).
Researched representations range from the full body \faUserMinus \cite{remelli2022drivable, cao2025real, luo2022artemis} to the upper body \faUserMinus \cite{tu2024tele} to the head \faUserMinus \cite{tran2024voodooArxiv, trevithick2023real} and hands \faUserMinus \cite{mundra2023livehand}.
While five out of six papers aim at a realistic representation of the human user, Luo et al. propose a system for rendering and animating interactive virtual animals \faUserMinus \cite{luo2022artemis}. At its core, their approach uses a differentiable neural representation tailored to model dynamic animals, including their fur, in real time.

\begin{table}[t]
\centering
    \caption{24 RF research contributing to \faUserPlus~XR-Study.}
    \label{XR-study}
    \small
    \begin{tabular}{lllll}
        \toprule
        Thematic topic         & Paper & Sub-category & Venue & OSS \\
        \midrule
        Interactive     &  \cite{sakashita2024sharednerf} & Collaborative systems & CHI '24    & y   \\
        experiences     &  \cite{huang2024virtualnexus}    & Collaborative systems & UIST '24 & y   \\
             &  \cite{patil2024radiance}     & Robot teleoperation & IROS '24 & y   \\
             &  \cite{li2024reality}     & Robot teleoperation & IROS '24 & y   \\
             &  \cite{kleinbeck2025multi}     & Medical applications & TVCG '25 & y   \\
        \midrule
        User-centric         & \cite{UCR_FoV_NeRF}     & Foveated rendering & TVCG '22 & y   \\
        rendering & \cite{UCR_VRS_NeRF}      & Foveated rendering & ISMAR '23 & y   \\
             &  \cite{URC_Scene_Aware_FoV_NeRF}     & Foveated rendering & TVCG '24 & y   \\
             &  \cite{UCR_Perceptual_RF_Foveated_Image_Synthesis}     & Foveated rendering& TVCG '24 & y   \\
             & \cite{Fan2025FovGSF3} & Foveated rendering& TVCG '25 & n \\
        \midrule
        Benchmarking              & \cite{Benchmarking_NeARPortation}      & Framework& VRST '22 & n   \\
             & \cite{Benchmarking_Fast_and_Robust_3DGS}      & Framework & SIGA-P '24 & y   \\
             & \cite{Benchmarking_Is3DGSUseful}      & User study & ISMAR '24 & n  \\
             & \cite{Benchmarking_creating_virtual_environments_3DGS}      & Pilot testing & VRW '25 &  n   \\
        \midrule
        Multi-modal              & \cite{multimodal_haptics_rendering_NeRF}      & Haptics rendering & UIST '23 & n   \\
        \midrule
        Content editing         & \cite{ren2024palettegaussian}      & Interactive color editing & ISMAR '24 & n   \\
             &  \cite{UCR_OmniPlane}     & Interactive color editing & TVCG '25 & y   \\
             & \cite{schieber2025semantics}      & Scene segmentation & VR '25 & y   \\
        \midrule
        Content gen.          & \cite{dai2025go}      & Context-aware text-to-3D & TVCG '25 & n   \\
        \midrule
        Spatial regist.          & \cite{zhai2025splatloc}      & Localization & TVCG '25 & n   \\
        \midrule
        Dynamic cap.          & na      & na & na & n   \\
        \midrule
        Optimization          & \cite{opt_kim2024superpixel}      &   Compression & VRST '24 & n   \\
        techniques     & \cite{opt_zhang2025srbf}      &   Compression & VR '25 & y   \\
             & \cite{opt_yin2024fsvfg}      & Compression & MM '24 & y   \\
             & \cite{opt_turki2024hybridnerf}      & Rendering & CVPR '24 & y   \\
        \bottomrule
    \end{tabular}
\end{table}

\begin{table}[t]
\centering
    \caption{42 RF research contributing to \faUserMinus~XR-Showcase.}
    \label{XR-Showcase}
    \small
    \setlength{\tabcolsep}{3.5pt}
    \begin{tabular}{lllll}
        \toprule
        Thematic topic         & Paper & Sub-category & Venue & OSS \\
        \midrule
        Interactive     &  \cite{li2023realitygit}     & Collaborative systems & ISMAR-A '23 & n   \\
        experiences     &  \cite{reynolds2024pop} & Collaborative systems & CHI-EA '24    & y   \\
             &  \cite{hu2024experiencing}     & Collaborative systems & UIST-A '24 & y   \\
             &  \cite{cao2025real}     & Avatars and agents & TVCG '25 & y   \\
             &  \cite{remelli2022drivable}     &  Avatars and agents & SIG '22 & y   \\
             &  \cite{luo2022artemis}     & Avatars and agents & TOG '22 & y   \\
             &  \cite{trevithick2023real}     & Avatars and agents & TOG '23 & n   \\
             &  \cite{tu2024tele}     & Avatars and agents & SIG '24 & y   \\
             &  \cite{tran2024voodooArxiv}     & Avatars and agents & TOG '24 & y   \\
             &  \cite{mundra2023livehand}     & Avatars and agents & CVPR '23 & y   \\
             &  \cite{li2024interactive}     & Industrial applications & VRW '24 & y   \\
             &  \cite{mcghee20243dcrewcap}     & Industrial applications & SIG-P '24 & y   \\
             &  \cite{zou2024arthronerf}     & Medical applications & ISMAR '24 & n   \\
             &  \cite{li2023interacting}     & XR interaction mechanisms & CHI-EA '23 & y   \\
             &  \cite{jiang2024vr}     & XR interaction mechanisms & SIG '24 & y   \\
        \midrule
        User-centric & \cite{UCR_Omnidirectional_NeRF}      & Omnidireaction  & VRW '22 & y   \\
        rendering     &  \cite{UCR_real_time_omnidireactional_roaming_large_scale_indoor_scenes}     & Omnidireaction & SIGA-TC '22 & y   \\
             &  \cite{UCR_Immersive_NeRF}     & Immersive light field & TVCG '23 & y   \\
        \midrule
        Benchmarking              & \cite{Benchmarking_immersive_ngp_poster}      & Toolkit & VRW '23 & n   \\
             & \cite{Benchmarking_VR_NeRF}      & Framework & SIGA '23 & y   \\
             & \cite{Benchmarking_explore_RF_content_generation_VR}      & Pilot testing & VRW '24 & y   \\
             & \cite{Benchmarking_pixilated_Choregography_VR}      & Pilot testing & VRW '25 & n   \\
        \midrule
        Multi-modal              & \cite{Multimodal_audio_3DGS}      & Audio & VRW '25 & n \\
                     & \cite{Multimodal_4D_haptics}      & Haptics rendering & SIG-IP '24 & y   \\
        \midrule
        Content editing     & \cite{ye2024AReal}      & Mesh insertion & TVCG '24 & y   \\
             & \cite{qiao2023dynamic}      & Mesh insertion & ICCV '23 & y   \\
            & \cite{waldowdimsplat}      & Diminished reality & VRW '25 & n   \\
             & \cite{aaa2025GaussianShopVR}      & Interactive editing techniques & VRW '25 & y   \\
             & \cite{mao2024live}      & Scene segmentation & VRW '25 & y   \\
             & \cite{wang2023neural}      & Scene disentanglement & CVPR '23 & y   \\
             & \cite{tran2024voodoo}      & Scene disentanglement & CVPR '24 & y   \\
        \midrule
        Content gen.          & \cite{vachha2024dreamcrafter}      & Speech-to-3D & UIST-A '24 & n   \\
        \midrule
        Spatial regist.          & \cite{park2023camp}      & Camera params. refinement & TOG '23 & n   \\
        \midrule
        Dynamic cap.          & \cite{aaa2025Realtime}      & Performance capture & VRW '25 & n   \\
             & \cite{zhao2022human}      & Performance capture & TOG '22 & y   \\
             & \cite{dong2023sailor}      & Performance capture & TOG '23 & y   \\
             & \cite{jiang2024robust}      & Performance capture & TOG '24 & y   \\
             & \cite{lombardi2021mixture}      & Facial capture & TOG '21 & y   \\
        \midrule
        Optimization          & \cite{opt_Dongye2025Human}      & Capture improvement  & VRW '25 & n   \\
        techniques     & \cite{opt_rojas2023re}      & Rendering & CVPR '23 & y   \\
             & \cite{opt_jiang2024hifi4g}      & Compression &   CVPR '24       & y   \\
             & \cite{opt_wang2024videorf}      & Compression & CVPR '24 & y   \\
        \bottomrule
    \end{tabular}
\end{table}

\textbf{Industrial and medical applications.} Six of the 19 \textit{Interactive Experience} contributions demonstrate the potential of RF for specific XR application fields.
These include robot teleoperation \faUserPlus \cite{li2024reality,patil2024radiance}, medical visualization \faUserMinus \cite{zou2024arthronerf}, \faUserPlus\cite{kleinbeck2025multi},
industrial facility inspection \faUserMinus \cite{li2024interactive},
and helicopter rescue crew training \faUserMinus \cite{mcghee20243dcrewcap}.

\textbf{XR interaction mechanisms}:
RF-scene and RF-object manipulations in an immersive environment can be performed via proxies.
Li et al. implemented a set of interactive features for NeRF models in a bounded volume, including exocentric manipulation, tunneling effects, and scene appearance editing \faUserMinus \cite{li2023interacting}.
Jiang et al. achieved physics-based deformations and dynamics of 3DGS objects by integrating mesh-based physics simulation into 3DGS \faUserMinus \cite{jiang2024vr}.

\textbf{XR integration considerations:}
Integration of RF representations into seamless interactive XR experiences faces various challenges, such as the interoperability of RF with the conventional render pipeline \cite{Benchmarking_immersive_ngp_poster}, compatibility with physics-based systems \cite{jiang2024vr}, and maintaining real-time performance during dynamic scene manipulation \cite{li2023interacting}.
It is noteworthy that many systems currently only support RF rendering or AR annotation overlays on screen space with synthesized depth images on XR displays for efficiency \cite{reynolds2024pop, sakashita2024sharednerf, patil2024radiance, zou2024arthronerf}.
Collisions and occlusions are also handled in screen space with rendered depth images \cite{sakashita2024sharednerf} or other modalities, such as scene mesh reconstruction \cite{reynolds2024pop}.
An RF rendering pipeline is typically integrated to XR systems either by (i) converting NeRF to meshes using marching cubes \cite{huang2024virtualnexus}, (ii) integrating native render plugins that bridge game engines like Unity with high-performance rendering backends (e.g., instant-ngp \cite{Benchmarking_immersive_ngp_poster, li2023realitygit, li2023interacting, li2024interactive}, the original CUDA-based 3DGS renderer \cite{li2024reality}, and a custom CUDA-based 3DGS physics simulator \cite{jiang2024vr}) that directly stream rendered outputs to the frame buffers of XR displays, or (iii) using an open-source integration of 3DGS in the Unity game engine with optimized parallel sorting capabilities for fast XR rendering \cite{kleinbeck2025multi}.
Applying super-resolution algorithms or deep learning super sampling (DLSS) is a popular technique for maintaining visual quality and high frame rates \cite{li2023realitygit, li2023interacting, li2024interactive, mundra2023livehand}.
In the supplementary material, we provide a detailed summary of XR integration techniques, interaction techniques, real-time performance, and hardware requirements of the surveyed papers in this \textit{Interactive Experience} category for supporting XR practitioners in designing, implementing, and evaluating interactive RF systems. 

\textbf{Hardware and performance}:
We observe a positive trend that RF rendering is becoming increasingly accessible, offering improved real-time performance and resolution with reduced computational demands. For example, Tran et al. demonstrated a NeRF-based VR telepresence system that achieved 30 fps stereoscopic rendering, albeit at a 512$\times$512 pixel resolution, and required two NVIDIA RTX 6000 GPUs for per-eye rendering \cite{tran2024voodoo}. Afterwards, Tu et al. presented a 3DGS telepresence system that runs at 30 fps at 2,048$\times$2,048 pixels on an autostereoscopic display with an NVIDIA RTX 4090 GPU \cite{tu2024tele}. Cao et al. demonstrated a 3DGS immersive telepresence system of holoported patients at 400 fps at 1080p with an NVIDIA RTX 6000 GPU \cite{cao2025real}.

\textbf{Challenges:}
i) In interactable RF, the relationship between scene objects might be dynamically changed by the user, which creates challenges regarding real-time lighting effects \cite{cao2025real, luo2022artemis, mundra2023livehand} as well as exposure of formerly occluded regions \cite{cao2025real,luo2022artemis,sakashita2024sharednerf}.
ii) Several of the considered interactive experiences rely on manual pre-processing steps, such as pre-defining a skeletal rig and capturing human motions \cite{luo2022artemis,remelli2022drivable} or setting object parameters \cite{jiang2024vr}.
iii) Interactive modifications of RF, e.g., through reenactment or ad-hoc object duplication, are computationally expensive, therefore limiting the use for standalone XR \cite{tran2024voodooArxiv,kleinbeck2025multi} or creating a tradeoff between rendering speed and visual fidelity \cite{sakashita2024sharednerf, huang2024virtualnexus}. 

\textbf{Opportunities:} i) Design and implement practical interactions that go beyond basic functionalities, such as changing the point of view or moving the scene \cite{luo2022artemis,kleinbeck2025multi,sakashita2024sharednerf,mcghee20243dcrewcap} (e.g., further investigations of simulating efficient physics-based interactions).
ii) Use of machine learning (e.g., large vision models) to monitor the plausibility of the RF scene after interactive changes (e.g., generated dynamics) \cite{jiang2024vr}. iii) Despite being the most directly applicable to XR, a large proportion of research in this category remains limited to XR-showcase, lacking both robust implementations and in-depth user studies. This gap presents significant opportunities for future work. iv) Integrate multi-modality in interactions (e.g. haptic and textile feedback), as described in Section \ref{subsec:multimodal}. v) Further investigations of applying hybrid representations (meshes + RFs) for supporting physics-based interactions, dynamic lighting, and real-time collision handling \cite{jiang2024vr}. 

\subsection{User-centric Rendering}
Contributions within user-centric rendering primarily focus on foveated rendering \cite{UCR_FoV_NeRF,UCR_VRS_NeRF,URC_Scene_Aware_FoV_NeRF, UCR_Perceptual_RF_Foveated_Image_Synthesis, Fan2025FovGSF3} and the generation of omnidirectional images or videos \cite{UCR_Omnidirectional_NeRF,UCR_real_time_omnidireactional_roaming_large_scale_indoor_scenes,UCR_OmniPlane}, aiming to render scenes based on human visual attention and perceptual relevance.

\textbf{Sampling regulation leveraging human visual acuity:} A key challenge in RF rendering, especially for scene representations requiring intensive neural network queries, is achieving high rendering speed for stereoscopic, high-resolution, and wide field-of-view displays on high-end HMD  \cite{UCR_FoV_NeRF}. As a result, early RF research within the XR community has focused on developing algorithms that leverage human visual perception characteristics to accelerate rendering, typically, via regulating sampling rates while maintaining comparable perceived rendering quality \cite{UCR_FoV_NeRF, UCR_VRS_NeRF, URC_Scene_Aware_FoV_NeRF, UCR_Perceptual_RF_Foveated_Image_Synthesis}.  Deng et al., first applied foveated rendering to NeRF in immersive VR settings. Their FoV-NeRF framework represents radiance fields using concentric spherical coordinates and employs multi-resolution image synthesis to exploit the fovea-peripheral characteristics of human vision \faUserPlus{\cite{URC_Scene_Aware_FoV_NeRF}}. To optimize stereoscopic rendering, they introduced a method that applies a disparity shift to peripheral images rather than generating separate images for both eyes. The introduction of instant-ngp \cite{Mller2022InstantNG} marked a significant advancement in terms of NeRF rendering speed by utilizing a hash encoding data structure to parallelize and accelerate neural network queries during ray casting \cite{Mller2022InstantNG}.  Building on this, Rolff et al. extended instant-ngp \faUserPlus{\cite{Mller2022InstantNG}} by incorporating variable rate shading (VRS) techniques, which further accelerated NeRF rendering by merging pixel blocks based on fovea-peripheral characteristics and saliency. In a similar line of work, Shi et al. designed a multi-ellipsoidal neural scene representation that retrains the network based on depth-based saliency to enhance rendering efficiency \faUserPlus{\cite{URC_Scene_Aware_FoV_NeRF}}, extending the FoV-NeRF framework. Likewise, Wang et al. proposed VPRF, a NeRF variant that regulates the sampling rate along each ray through a sampling weight-constrained training process, utilizing contrast-based saliency maps \faUserPlus{\cite{UCR_Perceptual_RF_Foveated_Image_Synthesis}}. Fan et al. integrated foveated rendering into a dynamic 3GDS pipeline by using a Gaussian forest representation, which separates dynamic and static components and selectively activates, deforms, and renders Gaussians based on visual acuity \faUserPlus\cite{Fan2025FovGSF3}.

\textbf{Bringing large-scale RF scenes to XR with outward-facing camera setup:} The original NeRF is typically reconstructed using perspective cameras with an inward-facing setup, which is optimized for accurately rendering a well-centered object. However, many immersive XR applications require building 3D environments with large-scale, unbounded scenes. In such cases, the traditional NeRF training pipeline often results in low-quality rendering and 
could cause unpleasant viewing experiences \cite{UCR_Immersive_NeRF}. To address this limitation, the CG and XR communities have begun investigating extensions of NeRF for omnidirectional rendering \cite{UCR_Omnidirectional_NeRF,UCR_real_time_omnidireactional_roaming_large_scale_indoor_scenes,UCR_OmniPlane} and immersive light fields \faUserMinus{\cite{UCR_Immersive_NeRF}}. For example, Li et al. introduced an approach to address uneven ray sampling in omnidirectional positional encoding by utilizing a Fibonacci sphere model which results in improved rendering quality \faUserMinus{\cite{UCR_Omnidirectional_NeRF}}. Huang et al. proposed a method to reconstruct omnidirectional RF from 360-degree image sequences using a geometry-aware RF network, enabling real-time 360-degree rendering of large indoor scenes \faUserMinus{\cite{UCR_real_time_omnidireactional_roaming_large_scale_indoor_scenes}}. For reconstructing large-scale, unbounded outdoor scenes, Yu et al. proposed a hybrid approach that combines background and foreground RF with different spatial mapping strategies for immersive light fields rendering based on RF representations \faUserMinus{\cite{UCR_Immersive_NeRF}}.

\textbf{Challenges:}
i) One major challenge in foveated rendering for NeRF lies in the use of multi-resolution image synthesis, which requires retraining multiple networks and inefficient storage consumption, further complicating the RF generation process \cite{UCR_FoV_NeRF}.
ii) Saliency-based user-centric rendering faces difficulties in generating accurate saliency maps in real time, particularly with complex or dynamic scenes \cite{UCR_Perceptual_RF_Foveated_Image_Synthesis, UCR_VRS_NeRF, UCR_Perceptual_RF_Foveated_Image_Synthesis,Fan2025FovGSF3}.
iii) The development of omnidirectional NeRF is still in its early stages, requiring more efficient representations and rendering methods to manage large-scale, full-sphere visual data \cite{UCR_Omnidirectional_NeRF, UCR_real_time_omnidireactional_roaming_large_scale_indoor_scenes}.
iv) All user-centric rendering methods currently encounter aliasing effects and rendering artifacts that could negatively impact user experiences \cite{UCR_Immersive_NeRF}.

\textbf{Opportunities:} i) Investigate more storage-efficient data structures and representations for training and RF generation, particularly when incorporating multi-modal saliency maps or synthesizing multi-image outputs. ii) Extend user-centric rendering methods to dynamic scenes, for instance, by predicting video saliency to accommodate motion and changing viewpoints. iii) Move beyond psychophysical experiments or simple demonstrations (e.g., 360 images) to evaluate user-centric rendering methods in real-world interactive XR environments.


\subsection{Benchmarking}\label{subsec:benchmarking}
The benchmarking theme category refers to contributions that extend existing methods with known techniques to achieve novel functionalities or insights into the RF representations, for example, by developing related toolkits to support accessibility \cite{Benchmarking_VR_NeRF,Benchmarking_Fast_and_Robust_3DGS, Benchmarking_NeARPortation, Benchmarking_immersive_ngp_poster} or performing user centric evaluation or pilot testing of XR integrations \cite{Benchmarking_Is3DGSUseful,Benchmarking_pixilated_Choregography_VR, Benchmarking_creating_virtual_environments_3DGS, Benchmarking_explore_RF_content_generation_VR}. 

\textbf{Frameworks and toolkits:} One important aspect of the benchmarking category includes contributions that extend existing RF methods by creating accessible toolkits or frameworks to facilitate easier usage of the RF techniques in XR. These contributions focus on building systems for RF capturing \cite{Benchmarking_VR_NeRF}, post-processing \cite{Benchmarking_Fast_and_Robust_3DGS}, remote streaming \cite{Benchmarking_NeARPortation}, and efficient multi-GPU rendering  \cite{Benchmarking_VR_NeRF, Benchmarking_Fast_and_Robust_3DGS}, utilizing known optimization techniques.
Hiroi et al. introduced NeARPortation, a system that enables real-time remote stereoscopic rendering of NeRF on an AR display. The system achieves full HD rendering at 35-40 frames per second (fps) by leveraging bidirectional client-server communication \faUserPlus \cite{Benchmarking_NeARPortation}.
Li et al. developed immersive-ngp, an open-source Unity toolkit that tackled the interoperability challenges of neural network-based RF models. It allows for rendering and visualizing instant-ngp \cite{Mller2022InstantNG} models in immersive VR environments by implementing a native C++ and CUDA plugin \faUserMinus \cite{Benchmarking_immersive_ngp_poster}. Xu et al. presented VR-NeRF, an open-source framework designed for capturing and rendering high-fidelity NeRF environments specifically for immersive VR experiences. VR-NeRF utilizes a perceptual color space for accurate HDR appearance and an efficient mip-mapping mechanism for level-of-detail rendering with anti-aliasing. It also further extends and enhances instant-ngp, achieving dual 2K HDR rendering at 36 Hz by utilizing a multi-GPU rendering setup \faUserMinus \cite{Benchmarking_VR_NeRF}.
Recently, Tu et al. introduced a fast and robust 3DGS rendering pipeline optimized for VR. This system employs mini-splatting techniques, incorporates a stop-the-pop (STP) solution, introduces optimal projection to mitigate ghosting effects, and integrates foveated rendering to enhance rendering speed \faUserPlus \cite{Benchmarking_Fast_and_Robust_3DGS}. The resulting pipeline delivers frame rates comparable to the original 3DGS implementation \cite{Kerbl20233DGS} with improved visual quality and user experiences \cite{Benchmarking_Fast_and_Robust_3DGS}.



\textbf{User evaluation and pilot testing:} Another line of work within the benchmarking category involves contributions that perform human-in-the-loop evaluations of photorealistic RF representations in XR. Kim et al. conducted a user study to investigate how different 3D visualization methods, including 3DGS, image-to-3D, and video playback, affect recognition memory and user experience. The study demonstrated that, in VR, 3DGS enables better object recognition memory \faUserPlus \cite{Benchmarking_Is3DGSUseful}. Qiu et al. introduced a pilot usability evaluation using the System Usability Scale (SUS) to compare the usability of mesh, 3DGS, and panoramic images for VR object creation. Their results revealed that 3DGS achieved a superior SUS score compared to meshed reconstruction and paranomic 3D scenes \faUserPlus \cite{Benchmarking_creating_virtual_environments_3DGS}. Additionally, there are several initial efforts focused on pilot testing the basic features of RF-based photorealistic scene representations for VR content creation \cite{Benchmarking_creating_virtual_environments_3DGS} \faUserMinus \cite{ Benchmarking_explore_RF_content_generation_VR}, as well as exploring artistic applications such as creating stop-motion animation in VR \faUserMinus \cite{Benchmarking_pixilated_Choregography_VR}.


\textbf{Metrics:}
Benchmarking RF in XR experiences often aims to strike a balance between frame rates (in frames per seconds, fps↑), latency (in milliseconds, ms↓), resolution (in the number of pixels, px↑, or  pixel per degree (PPD)↑ \cite{UCR_FoV_NeRF,li2024interactive}), and render quality (e.g., Structural Similarity Index Measure, SSIM↑,  Peak Signal-to-Noise Ratio, PSNR↑, and Learned Perceptual Image Patch Similarity, LPIPS↓). For remote streaming, it is also crucial to minimize round trip latency (RTL)↓ to below 20 ms \cite{Benchmarking_NeARPortation}. Evaluating foveated rendering algorithms for RF also employs psychophysical experiments to statistically test participants' subjective awareness of the peripheral degradation \cite{UCR_FoV_NeRF, UCR_VRS_NeRF, Fan2025FovGSF3}. As rendering artifacts could be amplified in immersive settings, causing discomfort or motion sickness, human-centric metrics such as the Simulator Sickness Questionnaire (SSQ) and SUS are used to holistically evaluate user experience \cite{Benchmarking_creating_virtual_environments_3DGS}.

\textbf{Challenges:} i) There is a lack of modular frameworks for benchmarking and exploring different variants of RF techniques in XR similar to Nerfstudio for the XR community to catch up with the rapid development in this research topic \cite{Tancik2023NerfstudioAM}.
ii) Neural network-based RF models still face compatibility challenges with traditional graphics pipelines, resulting in limited interactive and interoperation capabilities \cite{Benchmarking_immersive_ngp_poster}.
iii) For many RF variants, achieving high quality rendering still requires complex hardware setups, such as multi-GPU configurations for rendering \cite{Benchmarking_NeARPortation, Benchmarking_VR_NeRF} or camera rigs for high quality capturing, making existing framework less accessible for wider adaptation.
iv) RTL below 20 ms is desirable to minimize perceived negative effects from network latency when using remote neural rendering, which could be difficult to achieve depending on network situations \cite{Benchmarking_NeARPortation}. 
v) Existing human centric benchmarking efforts is still preliminary, focusing on simple exploratory testing and initial usability evaluation \cite{Benchmarking_creating_virtual_environments_3DGS,Benchmarking_Is3DGSUseful, Benchmarking_pixilated_Choregography_VR}. 


\textbf{Opportunities:}
i) Develop standardized frameworks to evaluate different variants of RF methods, similar to existing tools like NeRFStudio \cite{Tancik2023NerfstudioAM}, but focused on integrating popular XR platforms such as Unity and OpenXR for better accessibility for the XR community. 
ii) Standardize neural rendering models for game engines by compressing them into efficient formats (e.g., an ONNX model).
iii) Enhance single-shot capturing techniques and optimize RF rendering on mobile and low-resource devices for better accessibility in RF generation.
iv) Implement head and gaze motion prediction for remote neural rendering frameworks to mitigate perceived latency.
v) Conduct more in-depth user-centric benchmarks to provide deeper design insights for XR, including detailed comparison of textured mesh, NeRF, and 3DGS in XR and measuring perceived rendering quality, sense of presence, perceived cybersickness, etc.

\subsection{Multi-modal}\label{subsec:multimodal}

Multimodality in RF for XR refers to extending the rendering capabilities to additional sensory modalities, such as haptics \cite{multimodal_haptics_rendering_NeRF, Multimodal_4D_haptics} and audio \cite{Multimodal_audio_3DGS}. Current RF research primarily focus on visual realism, often neglecting other sensory dimensions. To enable truly immersive XR experiences, it is essential to develop 3D scene representations that not only achieve high visual fidelity but also provide congruence with other types of sensory feedback. As revealed in Tables \ref{XR-study} and \ref{XR-Showcase}, research in this category remains sparse. However, some pioneering contributions have begun to explore multi-modal RF. For instance, Zhang et al. developed the first 3 degrees of freedom (3DOF) haptic rendering method for NeRF, leveraging a stochastic haptic rendering technique to approximate and smoothen the underlying geometric information \faUserPlus \cite{multimodal_haptics_rendering_NeRF}. This approach demonstrates promising performance in mitigating the noisy collision response from implicit 3D representations \cite{multimodal_haptics_rendering_NeRF}. Additionally, Jiao et al. demonstrated an early prototype enabling interaction with free-viewpoint video augmented by haptic feedback, based on a hybrid NeRF variant that integrates explicit geometric structures with implicit visual appearance representations  \faUserMinus \cite{Multimodal_4D_haptics}.  Interestingly,  Yoshida et.al recently demonstrated the feasibility of extending 3DGS to novel-view acoustic synthesis by treating each pixel in the spectrogram as a point, and, similar to 3DGS, use spherical harmonics to represent view-dependent audio magnitude for continuous spatial audio representations \faUserMinus \cite{Multimodal_audio_3DGS}.


\textbf{Challenges:}
i) The lack of precise geometric information in NeRF, especially in regions reconstructed from sparse viewpoints or network hallucinations, can lead to inaccurate or very noisy haptics  collision responses \cite{multimodal_haptics_rendering_NeRF}.
ii) Haptic devices require high refresh rates to deliver realistic feedback \cite{multimodal_haptics_rendering_NeRF}, placing additional computational demands on the already expensive NeRF-based rendering process.
iii) Integrating multi-modal data—such as visual, audio, and haptic information—into a unified RF representation could introduce challenges in terms of storage efficiency, synchronization, and optimization workflows.

 \textbf{Opportunities:}
 i) Further extending RF rendering to additional multi-sensory modalities such as olfaction, thermal, and gustatory feedback, potentially by leveraging the efficiency of positional encoding to create continuous, high-fidelity representations from sparse data.
 ii) Incorporate material and textile-level feedback to enable more realistic haptics experiences.
 iii) Investigate haptics rendering with 3GDS, an area currently still unexplored, for example, by converting point based splats to distance fields \cite{multimodal_haptics_rendering_NeRF} or develop hybrid representations. 

\subsection{Content Editing}
Content editing in RF for XR includes contributions that modify RF representations, including appearance adjustments (e.g., recoloring, style transfer, and relighting) and scene manipulations (e.g., object insertion/removal for augmented/diminished reality).

\textbf{Object insertion and removal:}
Aiming to combine the strengths of RF and polygonal surface meshes, both \faUserMinus Qiao et al. \cite{qiao2023dynamic} and \faUserMinus Ye et al. \cite{ye2024AReal} propose hybrid approaches that embed meshes into a NeRF.
Both approaches estimate light sources in the NeRF scene and simulate realistic light transport from the NeRF to the embedded mesh and vice versa.
Qiao et al. additionally present a physics simulator that utilizes the SDF of the NeRF to detect and resolve collisions between the NeRF and meshes, including cloth, rigid and soft bodies \cite{qiao2023dynamic}.
Waldow et al. pursues the opposite goal of removing objects from an RF in real time, i.e. implementing diminished reality \faUserMinus \cite{waldowdimsplat}. The approach relies on Gaussian splats that are generated on a prior state of the scene and have to be manually aligned with the current real-world footage.

\textbf{Segmentation and decomposition of RF scenes:} Another line of research aims at building RF scene understanding and can precede further editing steps such as object insertion or color adjustment. 
Schieber et al. developed and empirically evaluated a method for semantics-controlled Gaussian splatting that segments the scene into classes \faUserPlus \cite{schieber2025semantics}.
Mao et al. prompt an LLM to analyze the physical properties of each scene object, thereby supporting subsequent physics-based interactions \faUserMinus \cite{mao2024live}.
Focusing on user interaction, Shen et al. showcase \textit{GaussianShopVR}, a platform that leverages a VR interface for 3DGS editing 
\faUserMinus \cite{aaa2025GaussianShopVR}.
Instead of segmenting an RF scene into different objects, Wang et al. disentangle the geometry and materials of a NeRF from lighting effects \faUserMinus \cite{wang2023neural}. This approach supports relighting, e.g. when meshes are inserted.
Tran et al. propose a technique for head reenactment that uses the appearance from a source image and the facial expressions from a driver video \faUserMinus \cite{tran2024voodoo}. Their proposed technique converts both the source and driver data into a shared 3D volumetric representation based on tri-planes.

\textbf{Interactive color editing:} Recently, Kou et al. extended omnidirectional NeRF to dynamic 360-degree videos, focusing on interactive content editing capabilities, such as recoloring \faUserPlus \cite{UCR_OmniPlane}. 
The method of Ren et al. also supports interactive color adjustment, both at the object and scene level \faUserPlus \cite{ren2024palettegaussian}. They propose three modes of interaction, namely manual, text-driven and image-driven, the latter being an example of style transfer to RF.

\textbf{Challenges:}
i) Current techniques for embedding mesh-based objects into RF scenes are limited in their ability to incorporate advanced lighting effects, such as indirect lighting, and to render shadows without artifacts \cite{qiao2023dynamic, ye2024AReal}.
ii) Two of the considered papers on scene segmentation and decomposition explicitly point out that they are currently limited to static scenes \cite{wang2023neural,schieber2025semantics}.
iii) Manual preparatory steps are required for several of the discussed approaches, such as pre-recording the scene for live object removal \cite{waldowdimsplat}, pre-defining labels for scene segmentation into classes \cite{schieber2025semantics}, and designing priors for geometry-lighting disentanglement \cite{wang2023neural}.

\textbf{Opportunities:}
i) Although being explicitly developed for XR, only four of the ten \textit{Content Editing} papers demonstrate their work on an immersive platform, such as a VR HMD \cite{schieber2025semantics, mao2024live, aaa2025GaussianShopVR, UCR_OmniPlane}. New interaction patterns for RF editing could emerge from the targeted use of such a platform, for example, through novel multimodal inputs in XR (e.g.combining hand gestures, gaze, and speech).

\subsection{Content Generation} \label{subsec:content_generation}

Content generation for RF in XR includes topics such as 3D scene authoring using generative AI models and is popular among the CG and CV community. As introduced in Section \ref{sect:trends}, while there are 28 contributions explicitly envisioned the applications of RF-based content generation to XR, only 2  have applied their methods in actual XR settings, indicating a significant research gap. Nonetheless, exsiting work already started to investigate deploying RF-based text-to-3D AI models in VR \cite{vachha2024dreamcrafter} and context aware generation of NeRF in a VR environment with consistent appearance \cite{dai2025go}. Vachha et.al presented Dreamcrafter, a framework that integrates generative AI algorithms for RF-based speech-to-3D content generation \faUserMinus \cite{vachha2024dreamcrafter}. They also introduced utilizing a proxy models (e.g. mesh) to compensate for the large generation latency \cite{vachha2024dreamcrafter}. Recently, Dai et.al introduced Go-NeRF, a novel framework for generating NeRF within an already established VR scenes based on text, user defined region, and the visual context of the scene’s surrounding. Go-NeRF distills 2D diffusion prior to 3D, thereby, creating content that could adhere to scene conditions (e.g. reflection and shawdow) \faUserPlus \cite{dai2025go}.


\textbf{Challenges:} i) Consistency between generated content and existing environments is essential for maintaining users' sense of immersion and presence in XR. This requires advances in realistic relighting, context-aware rendering, and geometry-consistent scene integration.
ii) Current RF-based text-to-3D generation methods have high latency, limiting their applicability in many real-time interactive XR scenarios. iii) Existing implementations have yet to achieve the level of physical realism needed for convincing blending for many MR experiences \cite{dai2025go}.

\textbf{Opportunities:} i). Further investigation in techniques for realistic inpainting and blending of heterogeneous 3D representations (e.g. seamless blending of a photorealitisc RF model with meshes with abstract visualization in augmented virtuality experiences). 
ii). Develop intuitive authoring interfaces for RF-based content creation that minimize cognitive load, which requires further investigation in methods for reducing response latency and improving accuracy.
iii) Conduct usability studies comparing the cognitive demands of different 3D content authoring methods (e.g. RF-based generation vs. traditional modeling methods). iv) Expand beyond controlled test datasets \cite{dai2025go} to real-world, immersive MR environments to investigate the practical applicability of existing methods.

\subsection{Spatial Registration}\label{subsec:spatial_regisration}
Spatial registration in RF is crucial for XR applications, as it enables real-time camera pose localization and the optimization  of camera parameters.
CamP by Park et al. \faUserPlus\cite{park2023camp} improves initial camera parameters (e.g., from a mobile device) by preconditioning the optimization in the NeRF training pipeline.
SplatLoc by Zhai et al. \faUserMinus\cite{zhai2025splatloc} leverages 3D Gaussian primitives to compress 3D scene models and introduces a salient 3D landmark selection algorithm to select a suitable subset of primitives for efficient camera localization.

\textbf{Challenges \& Opportunities:}
Similar to conventional discussions, key challenges remain in achieving robust, accurate, and energy-efficient registration in XR settings. This is especially true for mobile and head-worn systems operating in dynamic environments with limited compute budgets. Efficient scene representation, robust initialization and relocalization, and tight integration with real-time rendering pipelines are central to meeting the latency and visual coherence requirements of immersive applications.

\subsection{Dynamic Capture}
Capturing dynamic objects or humans is indispensable for representing users as photorealistic volumetric avatars in telepresence systems or rendering playback videos in XR applications. The key to achieving real-time rendering of dynamic human motion is an efficient RF representation, such as compression. Lombardi et al. \faUserMinus\cite{lombardi2021mixture} combine the advantages of volumetric and primitive-based approaches while minimizing computation in empty regions of space. Zhao et al. \faUserMinus\cite{zhao2022human} achieve neural surface reconstruction in minutes by bridging the traditional animated mesh workflow with RF rendering techniques. Jiang et al. \faUserMinus\cite{jiang2024robust} propose DualGS, which can compress human motion to integrate RF into XR environments seamlessly. Krause et al. \faUserMinus\cite{aaa2025Realtime} develop data storage and loading methods to achieve playback rates suitable for interactive VR experiences. For another stream of research, Dong et al. \faUserMinus\cite{dong2023sailor} present a generalizable method to handle unseen performers without fine-tuning, using very sparse RGBD live streams.

\textbf{Challenges \& Opportunities:}
While many RF-based methods for dynamic capture have been proposed in the CG community, their direct applications evaluated in XR settings remain limited. In fact, all contributions are categorized as \faUserMinus~XR-showcases.
This is due to the high computational cost associated with avatar rendering.
Moreover, existing approaches have not yet achieved the level of real-time high-fidelity rendering needed for compelling XR experiences.
Future work should focus on human-involved studies to determine whether reconstructed faces, bodies, and expressions in motion still fall within the uncanny valley for XR users.

\subsection{Optimization Techniques}

Unlike the benchmarking category, which focuses on building toolkits or framework to extend existing RF methods using known optimization techniques, a substantial amount of RF research in the CG, CV, MM, and robotics communities explores novel approaches to RF optimization. Optimization provides the foundation to RF's practicality in XR applications, and, as a result, this area encapsulate a wide range of research topics, including handling sparse input data, reconstructing large-scale scenes, improving capture quality \cite{opt_Dongye2025Human}, accelerating rendering speed \cite{opt_rojas2023re, opt_wang2024videorf, opt_zhang2025srbf}, and developing compression techniques \cite{opt_jiang2024hifi4g, opt_kim2024superpixel, opt_zhang2025srbf}. However,  
despite a total of 94 XR-envisioned contributions in this category, only 8 are XR-addressed, highlighting a significant gap between the envisioned application and real-world implementation. Nonetheless, exsting work already investigated in topics such as rendering on resource constrained devices (standalone XR headsets) \cite{opt_rojas2023re, opt_wang2024videorf}, compression for photorealistic performance streaming and remote rendering \cite{opt_jiang2024hifi4g, opt_kim2024superpixel, opt_zhang2025srbf}, and capturing improvement \cite{opt_Dongye2025Human}.

\textbf{Rendering on resource constrained devices:} Rojas et al. introduced Re-ReND, a real-time NeRF rendering framework for VR headsets that achieves 72 FPS by converting NeRF representations into a hybrid format: meshes, which are compatible with traditional graphics pipelines, and light fields, which could output pixel colors through a single-query \faUserMinus \cite{opt_rojas2023re}. In another line of work, Wang et al. proposed VideoRF, which enables 4D human performance rendering on VR headsets via remote streaming. Their method is based on a serialized 2D feature image stream, demonstrating the feasibility of real-time 4D NeRF playback in immersive settings \faUserMinus \cite{opt_wang2024videorf}.


\textbf{Compression:} Compression is a major challenge for 3DGS due to its high parameter storage requirements, especially for applications like 4D performance capturing and storage. To tackle this challenge, Jiang et al. proposed HiFi4G, a compact 4DGS representation for real-time human performance rendering \faUserMinus \cite{opt_jiang2024hifi4g}. Their method introduces a dual-graph mechanism—utilizing both fine-grain and coarse graphs to initialize Gaussians more efficiently \cite{opt_jiang2024hifi4g}. Kim et al. presented a superpixel-based initialization method for 3DGS, which extracts superpixels from keyframe images for sparse sampling, reducing the number of Gaussians \faUserPlus \cite{opt_kim2024superpixel}. They also conducted extensive benchmarks on memory consumption and evaluated the performance on VR devices. More recently, Zhang et al. introduced SRBF-Gaussian, a viewport-dependent color encoding technique using Spherical Radial Basis Functions (SRBFs) to enable selective transmission of viewport relevant color-data for more efficient 3DGS streaming and remote rendering \faUserPlus \cite{opt_zhang2025srbf}. Compression is also crucial for NeRF-based volumetric video streaming in telepresence and holoportation systems \cite{Mller2022InstantNG}. Yin et al. proposed adaptive feature grids, extending instant-ngp \cite{Mller2022InstantNG}, to enable level-of-detail (LoD) streaming via dynamic feature selection during raymarching. They evaluated its applicability for efficient volumetric video playback on VR devices \faUserPlus \cite{opt_yin2024fsvfg}.


\textbf{Capture Improvement:} Dongye et al. demonstrated the feasibility of human-in-the-loop 3DGS capture by enabling users to teleoperate a robotic arm in VR, such that viewpoints with limited viusal quality could be manually inspected and updated on demand  \faUserMinus \cite{opt_Dongye2025Human}. This approach also demonstrates the potential for continuous and adaptive 3DGS updates, which could further enhance 3DGS's application in remote collaboration systems \cite{li2023realitygit, sakashita2024sharednerf}. 

\textbf{Challenges:}
i) Optimization and training speed remain bottlenecks for real-time applications such as dynamic SLAM, spatial registration, and live capture and reconstruction \cite{opt_jiang2024hifi4g}.
ii) 3D/4DGS still relies on fast sorting operations, which limits scalability as the number of Gaussian increases \cite{opt_jiang2024hifi4g}.
iii) Approaches based on representation conversion (e.g., NeRF to mesh) could depend on accurate geometric information, presenting barriers in rendering fine-grained geometry \cite{opt_rojas2023re}.
iv) There is a speed-memory trade-off in optimization; faster rendering typically requires storing precomputed priors, increasing memory demands \cite{opt_wang2024videorf}. v) Accurate user behavior prediction for remote rendering remains an open challenge.

\textbf{Opportunities:}
i) Continue exploring novel representations to further extend the possibilities of optimal photorealistic scene representations. For example, 3DGS could be enhanced by incorporating scene-aware/semantic-aware reconstruction.
ii) The current optimization framework primarily focuses on accelerating rendering speed and compression. However, real-world XR implementations may require further investigation into optimizing performance during complex immersive interactions, such as physics-based interactions \cite{jiang2024vr} and haptic rendering \cite{multimodal_haptics_rendering_NeRF}.
iii) Many optimization techniques had already been developed prior to the milestones of 3DGS and instant-ngp ( e.g. hybridNeRF \cite{opt_turki2024hybridnerf}), making them valuable resources for inspiring future optimization strategies in other RF variants.
iv) Transfer and adapt the vast amount of available optimization techniques developed within other communities (CG, CV, and robotics) into actual XR setting. This could be achieved by encouraging more open source system \& toolkit contributions for RF in XR as mentioned in Section \ref{subsec:benchmarking}.
v) Continuous investigation in effective approaches for porting RF on mobile devices in terms of rendering speed and memory consumption. 

\section{Discussion}\label{sec:discussion}

\textbf{Limitations:} Our survey has several limitations. i) We focused on ICCV and CVPR to capture the current developments and research trends within the CV community. We could not include all major CV conferences, such as the European Conference on Computer Vision (ECCV), due to the large volume of papers.
Moreover, we did not include poster and workshop proceedings from CVPR and ICCV, as they are highly diverse and challenging to cover comprehensively within the scope of this survey. As a result, some emerging research trends from the CV community may not be fully represented.
ii) Given the vast number of papers, we focused on contributions related to XR directly or closely. As a result, some promising works had to be excluded, such as 360-degree scene representation contributions which had not extensively envisioned XR as core applications. Moreover, to maintain a structured and consistent presentation, we classified the papers according to the taxonomy and theme we developed. However, this categorization required assigning each paper to a single primary category based on its main contributions, even when some works could reasonably belong to multiple categories.
iii) We did not pursue a deep analysis of author identities. For example, we did not examine the overlap of authors between XR-Addressed and XR-Envisioned papers to investigate whether the same researchers contributed to both categories. An analysis in this direction could yield valuable insights.

\textbf{Ethical considerations:}
The ability of RF to synthesize photorealistic scenes, approaching a level of realism nearly indistinguishable from the real world, raises issues related to misinformation, identity misuse, and intellectual property (IP) violations. 
In immersive AR/MR environments, RF-generated content poses unique risks: users may struggle to distinguish between real and virtual elements.
At the same time, there is limited work on mechanisms like watermarking to protect creators' IP within RF-based representations.
Among all the XR-related papers we screened in full text, only one has focused on ethical-related topics in detail in embedding watermarks in NeRF \cite{Watermark_NeRF}.
This highlights a significant research gap in the intersection of RF and XR, one that demands urgent attention given the rapid pace of RF development and its integration into immersive systems.





\section{Conclusion}\label{conclusion}

In conclusion, we introduced the first survey paper dedicated to exploring the intersection of RF techniques and XR.
Through a systematic review, we identified key research gaps between how RF has been envisioned and how it has been implemented, particularly in areas such as content generation, spatial registration, optimization, and emerging topics like content editing and multimodal interaction (e.g. haptics).
We also point out the limited focus on ethical considerations despite the increasing realism and accessibility of RF-based content.
We believe this work will serve as an important resource for the XR community in understanding, navigating, and advancing the integration of RF technologies in XR.

\acknowledgments{
This work was supported by the Deutsche Forschungsgemeinschaft (DFG, German Research Foundation) under Germany's Excellence Strategy – EXC 2120/1 – 390831618, the European Union’s Horizon Europe research and innovation program under grant agreement No 101135025, PRESENCE project, and the MSCA Staff Exchanges project PLACES (grant agreement No 101086206). Early drafts of this submission were rephrased for clarity and grammatical correctness using OpenAI's GPT-4o model and Microsoft Copilot.}

\bibliographystyle{abbrv-doi}

\bibliography{template}



\section*{Supplementary material}

# Radiance Fields in XR: A Survey on How Radiance Fields are Envisioned and Addressed for XR Research
## – Supplementary Material –

Ke Li, Mana Masuda, Susanne Schmidt, Shohei Mori

## 1 PRISMA Flowchart

Fig. 1 shows the PRISMA flowchart [24], providing an overview of the data acquisition, screening, and eligibility process used in our survey. As showed in the figure, we started with identifying a large corpus of *RF-Mentioned* papers, papers that mention radiance fields. From the initial corpus of all *RF-Mentioned* papers, we performed an automatic and manual screening process to finally identify a collection of *XR-Mentioned* papers (papers only mentioned XR without specifically addressing an aspect of it) and *XR-Addressed* papers (papers addressed a crucial part of XR research or applications). On the selected papers, we performed an iterative tagging process to generate a taxonomy of RF for XR research applications, extending previously available RF taxonomies, which were specifically developed for the CV [5], CG [4], or robotic [37] communities.

## 2 A Practical Guide for Integrating RF into XR

Integrating RF into XR systems faces various challenges. To learn about practical solutions, this section reviews examples of XR systems from the surveyed papers in the *Interactive Experiences* theme (see Section 5.1 in the main paper).

### 2.1 Hardware and Performance

Table 1 provides an overview of the reviewed papers, detailing their application domains, the hardware or GPUs, running frame rates, display resolutions, types of display devices, and rendering engines.

The table shows a clear trend indicating that RF rendering has become more accessible with the introduction of more recent hardware and 3DGS in each application domain. Within the *Avatar/Agents* domain, the NeRF-based VR telepresence system achieved 30 fps in stereoscopic rendering at a 512×512 pixel resolution [34]. The system requires a pair of NVIDIA RTX 6000 GPUs, each of which is responsible for rendering content for one eye. In contrast, the 3DGS-based telepresence system equipped with an NVIDIA RTX 4090 runs at the same 30 fps at 2,024×2,024 pixels on an autostereoscopic display. A more recent 3DGS-based immersive telepresence system for patients' holoportation runs at 400 fps at 1080p resolution on an NVIDIA RTX 6000 GPU [2].

### 2.2 XR Pipeline with Mesh and RF

While we observed a rise in RF technologies, we also acknowledge limitations in comparison to conventional mesh rendering. We discuss

- Ke Li is with University of Hamburg, Germany.
  E-mail: ke.li@uni-hamburg.de
- Mana Masuda is with Keio University, Japan.
  E-mail: mana.smile@keio.jp
- Susanne Schmidt is with University of Canterbury, New Zealand.
  E-mail: susanne.schmidt@canterbury.ac.nz
- Shohei Mori is with University of Stuttgart, Germany.
  E-mail: s.mori.jp@ieee.org

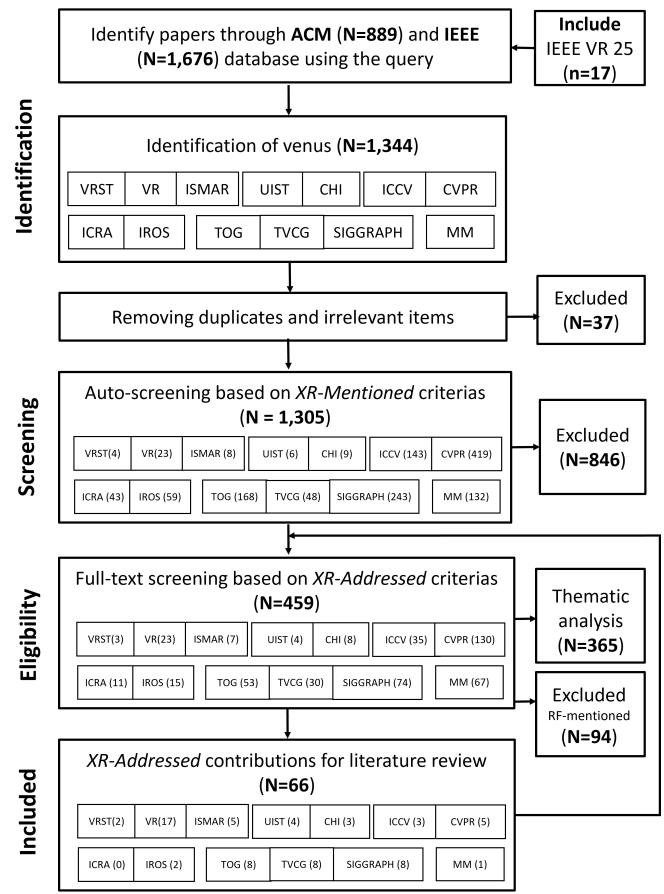

Fig. 1: PRISMA flowchart [24] giving an overview of the data acquisition, screening, and eligibility determination process. Here, numbers on SIGGRAPH include both *SIGGRAPH* and *SIGGRAPH Asia* conference proceedings.

this off-topic aspect in this supplemental document for XR practitioners and developers.

Table 2 overviews detailed comparisons of key components in the XR pipeline, including data preprocessing, modeling, rendering, and interaction. Typical tools for creating a textured mesh, NeRF, and 3DGS from photographs involve using photogrammetry software (e.g., Meshroom [1]), NeRF via instant-ngp [22], and 3DGS with its original CUDA implementation [11], respectively. For a comparison of pipelines for dynamic scenes, such as holoportation, we refer to the literature by Cao et al. [2].

**Data Preprocessing:** For data preprocessing, both textured mesh and RF methods require static high-resolution, high-quality multi-view images for sufficient viewpoint coverage. However, there is no defacto standard or sampling theories that support a fixed number of inputs and resolution for individual scenes for RF or photogrammetry [38]. The

Table 1: Overview of hardware requirements, render framerate, render resolution, and display types of all *Interactive Experiences* contributions.

| Year - Paper | Application Domain | Hardware | Framerate | Resolution | Display Type | Render Engine |
|---|---|---|---|---|---|---|
| 2023-Reality Git [16] | Collaborative System | RTX 3090 | unspecified | unspecified | VST-HMD | Native NeRF render plugin for Unity |
| 2024-SharedNeRF [30] | Collaborative System | unspecified | unspecified | 512 × 512 px | 2D Display | Native C++ and Direct3D renderer |
| 2024-Virtual Nexus [9] | Collaborative System | RTX 4090 | unspecified | n/a (mesh extraction from NeRF) | AR/VR HMD | unspecified |
| 2024-Pop-up Metaverse [28] | Collaborative System | n/a (cloud service) | > 60 fps | unspecified | 2D Display | WebGL and cloud service |
| 2024-Things2Reality [7] | Collaborative System | RTX 4070 | unspecified | unspecified | VR HMD | 3DGS Unity integration |
| 2024-RF for Robot Teleoperation [25] | Robot Teleoperation | RTX 4090 | 151 fps (3DGS), 0.9 fps (NeRF) | 1024 × 1024 px | VR/AR HMD (2.5D) | Remote streaming via TCP with Robotic Operation System (ROS) |
| 2024-Reality Fusion [15] | Robot Teleoperation | RTX 3090 | 30 − 35 fps (RF+point clouds) | 1536 × 1440 px per eye (727,019 Gaussians) | VR HMD | Native 3DGS CUDA render plugin |
| 2023-Interactive NeRF [17] | Interaction Mechanism | RTX 2080 | 20 − 30 fps | unspecified | VR HMD | Native NeRF render plugin for Unity |
| 2024-VR-GS [10] | Interaction Mechanism | RTX 4090 | 30 − 120 fps (depends on vertex and number of gaussian) | varied (depends on meshes and number of gaussian) | VR HMD | Native CUDA plugin for Unity |
| 2022-Volumetric Avatar [27] | Avatars/Agents | unspecified | 60 fps | 1024 × 666 px | 2D display | unspecified |
| 2022-Artemis [20] | Animated Animals | TITAN RTX / RTX 3090 | 29.16 / 44.07 fps | 512 × 512 × 512 | VR HMD | unspecified |
| 2023-Single-Image Portrait [35] | Avatars/Agents | RTX 4090 | 24 fps | unspecified | 2D Display | unspecified |
| 2023-LiveHand [23] | Avatars/Agents | RTX 3090 | 45 fps | 512 × 334 px | 2D Display | Custom NeRF renderer |
| 2024-VoodooXR [33] | Avatars/Agents | 2 x RTX 6000 | 30 fps | 512 × 512 px | VR HMD | Custom integration with Unity |
| 2024-Tele-Aloha [36] | Avatars/Agents | RTX 4090 | 30 fps | 2048 × 2048 px | autostereoscopic display | generalizable custom 3DGS renderer |
| 2025-Holographic Patient [2] | Avatars/Agents | RTX 6000 | 400 fps | 1080 px | Hololens 2 + PC | Custom 3DGS shader in Unity |
| 2024-magicNeRFlens [18] | Virtual Facility Inspection | RTX 3090 | 30 fps | 20 PPD & 30 FoV | VR HMD | Native NeRF render plugin for Unity |
| 2024-Helicopter [21] | Rescue Crew Training | RTX 6000 | unspecified | unspecified | VST-HMD | unspecified |
| 2024-ArthroNeRF [41] | Medical Visualization | RTX 3080 | 22.4 fps | unspecified | AR HMD (2.5D) | unspecified |
| 2025-Multi-layer 3DGS [13] | Medical Visualization | RTX 4090 | > 100 fps | 2064 × 2208 px per eye | VR HMD | 3DGS Unity integration [26] |

Table 2: End-to-end comparison of pipeline for reconstructing, modeling, rendering, and interacting with mesh (generated via photogrammetry), 3DGS, and NeRF in XR.

| Pipeline Stage | Aspect | Textured Mesh (Photogrammetry with Meshroom [1]) | 3DGS (Kerbl et al. [11]) | NeRF (instant-ngp [22]) |
|---|---|---|---|---|
| **Data Preprocessing** | Input Images | Varies from 20 to hundreds, depending on scene size and application | Varies from 20 to hundreds, depending on scene size and application | Varies from 20 to 200, depending on scene size and application |
| | Image Resolution Requirements | Static high resolution | Static high resolution | Static high resolution |
| | Preprocessing Pipeline | SfM → MVS → Dense point cloud | SfM → Sparse point cloud | SfM → Camera instrinsics &poses |
| **3D Modeling** | Representation | Explicit mesh (vertices, faces, UV) | Explicit 3D Gaussian "point" (incl. poses, scale, opacity, color) | Implicit MLP (scene function) |
| | Processing Pipeline | Complex: meshing + texturing + manual post processing | **Simple: 3DGS optimization (2-5 minutes with CUDA)** | **Simple: network training (2-5 minutes with CUDA)** |
| | Photorealism | Medium–High (depends on texture quality and complexity of scene details) | **High- very high given enough view coverage** | **High–very high given enough view coverage** |
| | Editability | **High (possible using standard 3D software, e.g. Blender)** | Medium (splat-level editing) | Low (updating network typically necessary) |
| **Rendering** | Rendering Method | Rasterization | Point sorting and blending | Volumetric ray casting |
| | XR Rendering Performance | **Mostly real-time (depends on number of vertices and texture resolution)** | **Mostly real-time (depends on number of Gaussians and sorting efficiency)** | Real-time performance possible with foveated rendering or other techniques in optimizing the number of rays and samples |
| **XR Integration & Interaction** | Game Engine Integration | Fully supported (Unity, Unreal, Blender, etc) | Possible (e.g. via custom shaders in Unity [26], Unreal [8], or Blender [6]) | Not directly supported, but possible via native render plugin [40] |
| | Export Format | Standard .obj, .fbx, .glb | Custom point-based file (e.g. ply [26]); | binary MLP checkpoint file |
| | Interactivity | **High, compatible with standard physics simulation and collision handling methods** | Medium, with some geometric primitives available for physics-based simulation and collision handling using hybrid representation [10] | Low, due to implicit surface |

current findings in the CV and CG areas suggest that the number of required images depends on scene complexity, typically ranging from 20 to 100 for single-object reconstructions [17] and several hundred for large-scale environments [17]. Moreover, Kopanas et. al suggested taking images as uniformly as possible to achieve the best synthesis results [14]. Otherwise, the so-called next-best-view sampling approaches are possible solutions, but this involves RF reconstruction every time a new view is inserted [32]. Despite similar input requirements, their preprocessing pipelines differ. In mesh-based pipelines, structure-from-motion (SfM) [31] is used to initialize multi-view stereo (MVS) for dense point cloud reconstruction. For NeRF, SfM is used to estimate camera intrinsics and poses required to train the coordinate-based multilayer perceptron (MLP). In 3DGS, SfM generates sparse point clouds to initialize Gaussian optimization.

**3D Modeling:** Compared to textured mesh pipelines, modeling NeRF and 3DGS requires significantly less manual effort. NeRF and 3DGS workflows involve only network training or gradient descent optimization out of the box. Textured mesh creation through photogrammetry consists of multiple stages, such as mesh extraction, surface texturing, and sometimes manual post-processing to correct artifacts and texture misalignment [19].

Additionally, NeRF and 3DGS provide superior photorealism by modeling view-dependent appearance effects, such as specular highlights and reflections, through learned or optimized opacity in the radiance fields. Meshes produced with photogrammetry, however, typically use back-projected textures, which fail to capture such view-dependent variations. As a result, empirical user studies [12] also report that RF-based models are perceived as having significantly higher visual quality than textured meshes. Moreover, both NeRF and 3DGS can achieve high visual fidelity even in complex scenes (PSNR $> 25$ and SSIM $> 0.8$ [11, 22]) with fine details such as fur and hair [22], which textured meshes often fall short.

However, a key limitation of NeRF and 3DGS lies in their editability. There are many well-established tools like Blender for textured meshes, but tools for NeRF and 3DGS are still evolving. Editing NeRF or 3DGS typically requires retraining or re-optimizing the model, for example, by incorporating semantic segmentation to enable object-level modifications [39].

**Rendering:** Textured meshes, NeRF, and 3DGS differ in rendering methods. Textured meshes leverage efficient hardware-accelerated rasterization. NeRF relies on expensive volumetric ray casting. 3DGS offers a compromise with point-based rendering that enables real-time performance and high visual quality. Although the original 3DGS rendering pipeline is implemented using CUDA, there already exist some open projects implementing 3DGS using graphics API (e.g., WebGL[1] and Vulkan[2]).

XR rendering performance for all methods depends on scene com-

---
[1] https://github.com/antimatter15/splat
[2] NVIDIA Developer Technical Blog, "Real-Time GPU-Accelerated Gaussian Splatting with NVIDIA DesignWorks Sample vk_gaussian_splatting"

plexity. Textured meshes are typically real-time unless the scene includes a large number of vertices or ultra-high-resolution textures. 3DGS performance varies with the number of Gaussians and sorting efficiency. NeRF rendering is typically slower than 3DGS and textured meshes but can achieve real-time performance in XR with optimizations like foveated rendering [3, 29] and field-of-view restriction [18], which reduce neural network queries by limiting rays and samples.

**XR Integration & Interaction:** In terms of XR integration and interaction, textured meshes are fully supported in standard game engines and tools such as Unity, Unreal, and Blender. They use widely adopted formats like `.obj`, `.fbx`, or `.glb`, and support high interactivity, being compatible with standard physics simulation and collision handling systems. 3DGS integration is possible through custom point-based shaders and file formats (e.g., `.ply`) in engines like Unity[3] and Unreal[4]. Interaction capabilities are moderate, with some support for physics and collision handling through hybrid representations and custom physics simulators [10]. NeRF, due to its implicit nature, lacks direct support in game engines. Integration is only possible via native rendering plugins that stream the outputs directly to the HMD, and interactive capabilities remain limited due to the absence of an explicit surface geometry.

**Summary & Recommendations:** In summary, the key advantage of RFs lies in their superior photorealistic rendering capabilities, particularly in synthesizing fine details and realistic view-dependent lighting of real-world scenes, along with their low-effort modeling pipeline. However, compared to mesh-based representations, RFs still face notable limitations in terms of editability, interactivity, and standardized integration into common game engines and graphics tools. Nonetheless, we expect more solutions will arise to mitigate these challenges in the near future, given the exponential growth of research in the field. As reflected in the *Interactive Experiences* category (see Section 5.1 of the main paper), current RF techniques are well-suited for interactive XR applications, such as remote collaboration, avatar-based communication, robot teleoperation, and medical scenarios, where quick modeling of real-world scenes with high visual fidelity is essential. In contrast, textured meshes are generally more suitable for constructing virtual environments and assets that demand higher levels of interactivity and customization.

## ACKNOWLEDGMENTS

---

[3] https://github.com/aras-p/UnityGaussianSplatting
[4] https://github.com/xverse-engine/XV3DGS-UEPlugin


\end{document}